\documentclass[%
reprint,
 amsmath,amssymb,
 aps,superscriptaddress,
pra,
]{revtex4-1}

\usepackage{dcolumn}
\usepackage[T1]{fontenc}
\usepackage{siunitx}

\usepackage[english]{babel}
\usepackage[utf8]{inputenc}
\usepackage{amsmath}
\usepackage{graphicx}
\usepackage[colorinlistoftodos]{todonotes}

\usepackage{hyperref}
\usepackage{bm}

\newcommand{\p}{\partial}

\newcommand{\bnabla}{\bm{\nabla}}
\newcommand{\ex}{\bm{\hat{e}}_1}

\newcommand{\ez}{\bm{\hat{e}}_3}
\newcommand{\emn}{\epsilon_{\mu\nu}}

\newcommand{\magn}{\bm{m}}

\newcommand{\DM}{D}
\newcommand{\dm}{\lambda}

\newcommand{\qfactor}{\kappa}

\newcommand{\storque}{\beta}
\newcommand{\STorque}{B}
\newcommand{\Je}{J}
\newcommand{\Jp}{J_0}

\newcommand{\lex}{\ell_{\rm ex}}
\newcommand{\ldm}{\ell_{\rm D}}

\newcommand{\Hext}{H_{\rm ext}}

\newcommand{\bHext}{\bm{H}_{\rm ext}}
\newcommand{\bhext}{\bm{h}_{\rm ext}}
\newcommand{\bfext}{\bm{f}_{\rm ext}}
\newcommand{\bHm}{\bm{H}_{\rm m}}
\newcommand{\bhm}{\bm{h}_{\rm m}}

\newcommand{\tmagn}{\mathcal{M}}
\newcommand{\vel}{\mathit{v}}

\newcommand{\skyrmion}{q}
\newcommand{\Skyrmion}{Q}

\begin{document}
\preprint{APS/123-QED}
 
\title{Chiral skyrmion auto-oscillations in a ferromagnet under spin transfer torque}
\author{Naveen Sisodia}
\affiliation{Department of Physics, Indian Institute of Technology Delhi, Hauz Khas, New Delhi 110016, India}
\author{Stavros Komineas}
\affiliation{Department of Mathematics and Applied Mathematics, University of Crete, 70013 Heraklion, Crete, Greece}
\author{Pranaba Kishor Muduli}
\affiliation{Department of Physics, Indian Institute of Technology Delhi, Hauz Khas, New Delhi 110016, India}
\date{\today}

\begin{abstract}
A skyrmion can be stabilized in a nanodisc geometry in a ferromagnetic material with Dzyaloshinskii-Moriya (DM) interaction.
We apply spin torque uniform in space and time and observe numerically that the skyrmion is set in steady rotational motion around a point off the nanodisc center.
We give a theoretical description of the emerging auto-oscillation dynamics based on the coupling of the rotational motion to the breathing mode of the skyrmion and to the associated oscillations of the in-plane magnetization. 
The analysis shows that the achievement of auto-oscillations in this simple set-up is due to the chiral symmetry breaking.
Thus, we argue that the system is turned into a spin-torque oscillator due to the chiral DM interaction.
\end{abstract}


\maketitle

\section{Introduction}
Auto-oscillations of magnetization is among the primary phenomena which are responsible for the wide range of applications of magnetic materials.
The most common method for achieving auto-oscillations at the nanoscale is by using the phenomenon of spin transfer toque (STT) \cite{berger1996emission,slonczewski1996current}.
It has been shown that spin-torques may induce precessional motion of uniform magnetization \cite{kiselev2003microwave,RippardPufallKaka_PRL2004,chen2016spin}
or rotation of vortices \cite{ivanov2007excitation,pribiag2007magnetic,FinocchioOzatay_PRB2008,khvalkovskiy2009vortex,dussaux2010large,Komineas_EPL2012}.
Magnetization auto-oscillations can be used to build spin-torque nano-oscillators (STNO) \cite{kiselev2003microwave,deac2008np,vincent2009ieeejssc,silva2008developments,kim2012spin,KATINE20081217,RussekRippardCecil_2010,chen2016spin,sharma2017high,muduli2012decoherence,sani2013mutually,mohseni2013spin} that promise to be the core parts of frequency-generator devices at the nanoscopic scale for microwave communication.

Magnetic skyrmions are topological nanoscale magnetization configurations which are stabilized in materials with the Dzyaloshinskii-Moriya interaction (DMI) \cite{BogdanovHubert_JMMM1994,BogdanovHubert_JMMM1999}.
The dynamics of skyrmions was recently proposed to be exploited for an STNO, as this can be expected to offer nearly two orders of magnitude lower threshold current than conventional STNOs \cite{ZhangWang_NJP2015,GarciaSanchezSampaio_NJP2016}.
STNOs based on localized excitations could present high output power \cite{dussaux2010large} and narrow spectral linewidth \cite{pribiag2007magnetic}.
Skyrmions are strictly localized objects, unlike vortices that have a small core but their configuration is extended over the whole film.
In previous studies, oscillations of skyrmions were established by using a non-uniform spin torque, by either employing a vortex state reference layer to obtain spatially varying polarization~\cite{GarciaSanchezSampaio_NJP2016} or by using a nano-contact geometry with spatially varying current density~\cite{ZhangWang_NJP2015}.

Here, we demonstrate numerically and describe theoretically magnetization auto-oscillations due to sustained rotation of the skyrmion, by using a simpler system which is easier to realize.
Our system consists of a disc-shaped element with perpendicular anisotropy and interfacial DMI.
We assume {\it time independent} and {\it spatially uniform} spin-torque across the magnetic layer.
We show that steady-state auto-oscillations in the form of rotational motion of the skyrmion in the disc are possible when these get coupled to the breathing mode of the chiral skyrmion \cite{SchuetteGarst_PRB2014}.
The detailed theoretical description is based on phenomena that arise due to the chiral symmetry breaking.
Thus, the studied auto-oscillations depend crucially on  the presence of the DMI.

The present set-up may be the simplest that has been proposed to-date in order to produce oscillations based on a localized magnetic soliton.
This makes the system interesting for experimental realization and also quite appealing for a detailed theoretical study.

Our simulations and theoretical analysis describe only the ferromagnetic free layer. The proposed system can be realized in any system where uniform spin torque can be generated.
In the following, we will assume that the system is realized as a magnetic tunnel junction (MTJ) where a spin transfer torque originates from a spatially uniform \textit{dc} current from a uniformly magnetized reference layer.
However, the system can also be realized as a spin-Hall nano-oscillator \cite{chen2016spin,demidov2012magnetic} using the spin current generated from a heavy metal.

\section{Skyrmion in a disc}
\label{sec:statcSkyrmion}

We consider a disc-shaped magnetic element of a material with DMI originating in the interface with a heavy-metal layer.
Micromagnetics code $mumax^3$~\cite{vansteenkiste2014design} was used to numerically simulate the skyrmionic magnetization texture in the disc.
We use values for the material parameters similar to those measured for the $\textrm{Co}_2\textrm{FeAl}$ Heusler alloy~\cite{CoFeAl_arxiv_2017}, as shown in Table~\ref{table:mat_parameter}.
But, the DM constant is chosen to be higher than in Ref.~\cite{CoFeAl_arxiv_2017} in anticipation of improvements in the material fabrication.
\begin{table}[h]
\centering 
\begin{tabular}{c c}
\hline
Parameter & Value \\
\hline
$M_s$ & $838\times 10^3\,{\rm A/m}$ \\
A & $11\times 10^{-12}\,{\rm J/m}$ \\
K & $3\times 10^5\,{\rm J/m^3}$ \\
D & $1.7\times 10^{-3}\,{\rm J/m^2}$ \\
\hline
\end{tabular}
\caption{The values for material parameters used in our standard simulation: $M_s$ is the saturation magnetization, $A$ the exchange constant, $K$ the easy-axis anisotropy constant, $D$ the interfacial DMI constant.}
\label{table:mat_parameter}
\end{table}
The disc has diameter $d=150\,\textrm{nm}$ and thickness $d_f=5\,\textrm{nm}$.
We discretized the entire geometry in cuboidal cells each having a size of $0.5\,\textrm{nm}\times 0.5\,\textrm{nm} \times 5\,\textrm{nm}$ along the $x, y$ and $z$ directions respectively.
The change in magnetization along the $z$ direction (perpendicular to the disc) is assumed to be negligible.

\begin{figure}[t]
\begin{center}
\includegraphics[width=\columnwidth]{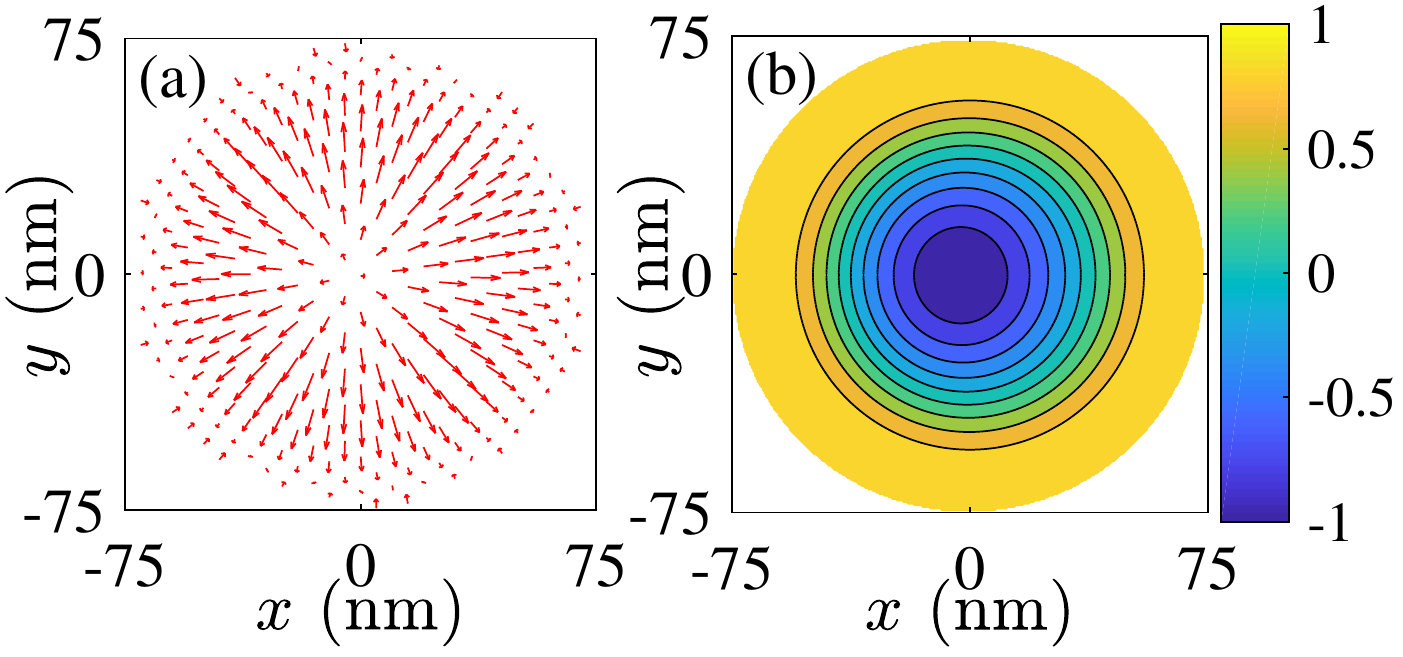} 
\caption{(a) Vector plot for a skyrmion in the center of a nanodisc with diameter $d=150\,{\rm nm}$ and thickness $d_f=5\,{\rm nm}$.
The projection of the magnetization on the plane of the disc is shown.
This N\`eel-type skyrmion is a static magnetization configuration, for the parameter values in Table \ref{table:mat_parameter} under a $\Hext=0.1~{\rm T}$ external field.
(b) Contour plot for the out-of-plane component of the magnetization ($m_3$) for the same skyrmion.
}
\label{fig:staticSkyrmion}
\end{center}
\end{figure}

We start a simulation applying a high out-of-plane field (1~T) which aligns the magnetization along the negative $z-$direction.
We then remove the external field and we observe that the magnetization relaxes to a skyrmion configuration.
Once the skyrmion is formed, a small uniform out-of-plane field $\Hext$ along the positive $z$ direction is applied in order to tune (reduce) the size of the skyrmion.
Fig.~\ref{fig:staticSkyrmion} shows the static skyrmion obtained for $\Hext=0.1~{\rm T}$.
Our numerical result shows that the skyrmion can be easily obtained and is a stable state for the present system, although it is not the ground state.
A stable skyrmion is obtained by the above procedure for the range of DM parameter values $\DM=1.5 - 1.8\,{\rm mJ/m^2}$.

The parameter values in Table~\ref{table:mat_parameter} give the dimensionless anisotropy constant (quality factor)
\begin{equation}
\qfactor = \frac{2K}{\mu_0 M_s^2} = 0.68.
\end{equation}
Such a value for the perpendicular anisotropy would normally be fully compensated by the magnetostatic field and force the magnetization to lie in the film plane. 
This value of $\qfactor$ is lower compared to the value used in Ref.~\cite{ZhangWang_NJP2015,GarciaSanchezSampaio_NJP2016} where the nanodisc was perpendicularly magnetized ($\qfactor>1$). 
Similar calculations have been done for magnetic bubbles (that are also skyrmionic textures) in magnetic elements with perpendicular anisotropy.
It was found that these can be stabilized in confined geometries \cite{DruyvesteynSzymczak_PSSA1972,IgnatchenkoMironov_JMMM1993,Komineas_PRB2005}
and they have been observed in experiments \cite{MoutafisKomineas_PRB2007}.
They can also be stable in materials with $\qfactor < 1$ despite the effective easy-plane anisotropy in this case \cite{MoutafisKomineas_PRB2006}.
While magnetic bubbles exist in infinite films only when an external bias field is applied, they are stabilized in dots without an external field thanks to the magnetostatic field from the surfaces of the magnetic element.
DMI further stabilizes skyrmionic textures in a disc \cite{LeonovMostovoy_EPJ2013,BegStampsFangohr_srep2015}. 

Every magnetic configuration is characterized by the  skyrmion number, defined as
\begin{equation}  \label{eq:skyrmionNumber}
\Skyrmion = \frac{1}{4\pi}\int \skyrmion\, dx dy,\quad \skyrmion = \emn \bm{m}\cdot (\p_\nu\bm{m}\times \p_\mu\bm{m})
\end{equation}
where $\magn$ is the magnetization normalized to $M_s$, $\emn$ is the totally antisymmetric tensor with $\mu,\nu=1,2$, and spatial derivatives are denoted by $\p_\mu$ with $\p_1=\p_x, \p_2=\p_y$.
It takes integer values only, and gives $\Skyrmion=1$ for skyrmions such as those studied in the present work, if the integration in Eq.~\eqref{eq:skyrmionNumber} would extend to spatial infinity.
On the other hand, in the confined geometry of the nanodisc we find $\Skyrmion = 0.989$ for the static skyrmion in Fig.~\ref{fig:staticSkyrmion}.
We numerically calculate $\Skyrmion$ by summing contributions of the topological density $\skyrmion$ up to a distance of $8~\textrm{nm}$ from the boundary of the nanodisc. 
We thus exclude from the calculation the magnetic configuration at the boundary that may be far from uniform due to the boundary conditions implied by the DM interaction.

\section{Skyrmion rotation}

We start a simulation with the skyrmion shown in Fig.~\ref{fig:staticSkyrmion}, and we apply a current that is constant in time, has uniform density across the disc and it is polarized in the in-plane $\ex$ direction (\textit{x}-direction).
We choose a current $I=2.0\,\textrm{mA}$ that corresponds to a current density $\Je=113\,{\rm GA/m^2}$.
The polarization efficiency is taken to be $b=0.54$ while the 
Gilbert damping is taken to be $\alpha=0.01$.

\begin{figure}[htb]
\begin{center}
\includegraphics[width=\columnwidth]{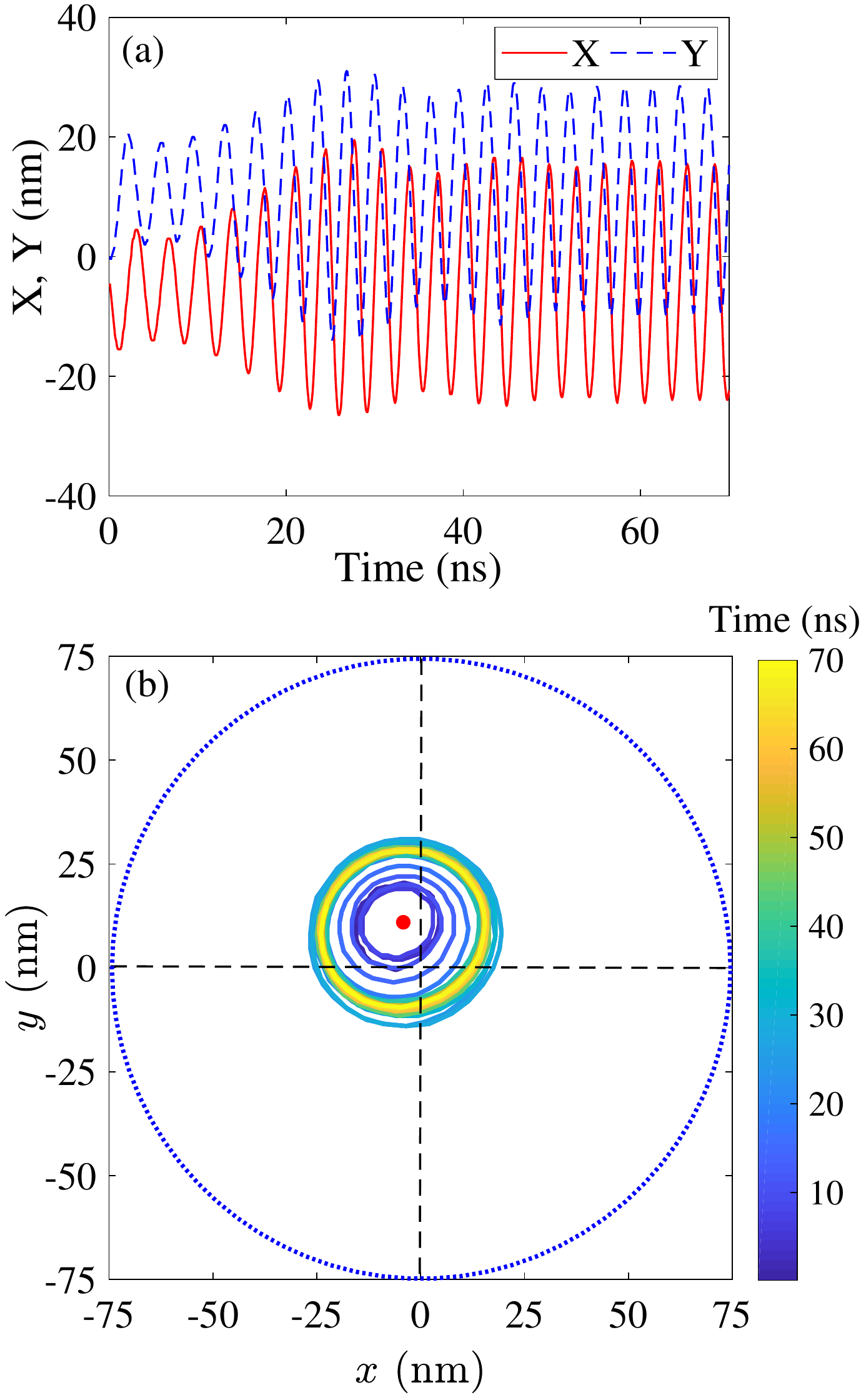}
\caption{(a) Skyrmion position coordinates $(X,Y)$ as functions of time under the influence of spin torque. (b) Trajectory of the skyrmion in the nanodisc. The skyrmion is assumed to be at the point where $m_3=-1$. It is initially located at the nanodisc center, and the current is switched on at time $t=0\,\textrm{ns}$. Steady-state periodic motion is established after $t\sim 40\,\textrm{ns}$. The red dot in (b) shows the center of the sustained rotational motion, located at $(-4.2\,\textrm{nm}, 11.0\,\textrm{nm})$ and the outer blue circle shows the nanodisc boundary. The colorcode shows time.
}
\label{fig:XY}
\end{center}
\end{figure}


We follow the skyrmion position $(X,Y)$ by tracing the point where $m_3 = -1$.
This definition of the skyrmion position is to a certain extent arbitrary and might not always reflect the mean skyrmion position.
Fig.~\ref{fig:XY}(a) shows the position coordinates as functions of time starting at time $t=0$ when the current is switched on, and for several cycles during the auto-oscillation motion.
Fig.~\ref{fig:XY}(b) shows the corresponding trajectory of the skyrmion position.
Spin-torque displaces the skyrmion from the center of the nanodisc, and this moves initially to the left.
The trajectory is then curved due to the force from the boundary and, eventually, a steady-state rotation is established. 
The trajectory is slightly elliptical.
In our full simulation, we have followed the motion for about one thousand cycles obtaining clear numerical evidence that there is no decay of the rotational motion.

\begin{figure}[t]
\begin{center}
\includegraphics[width=\columnwidth]{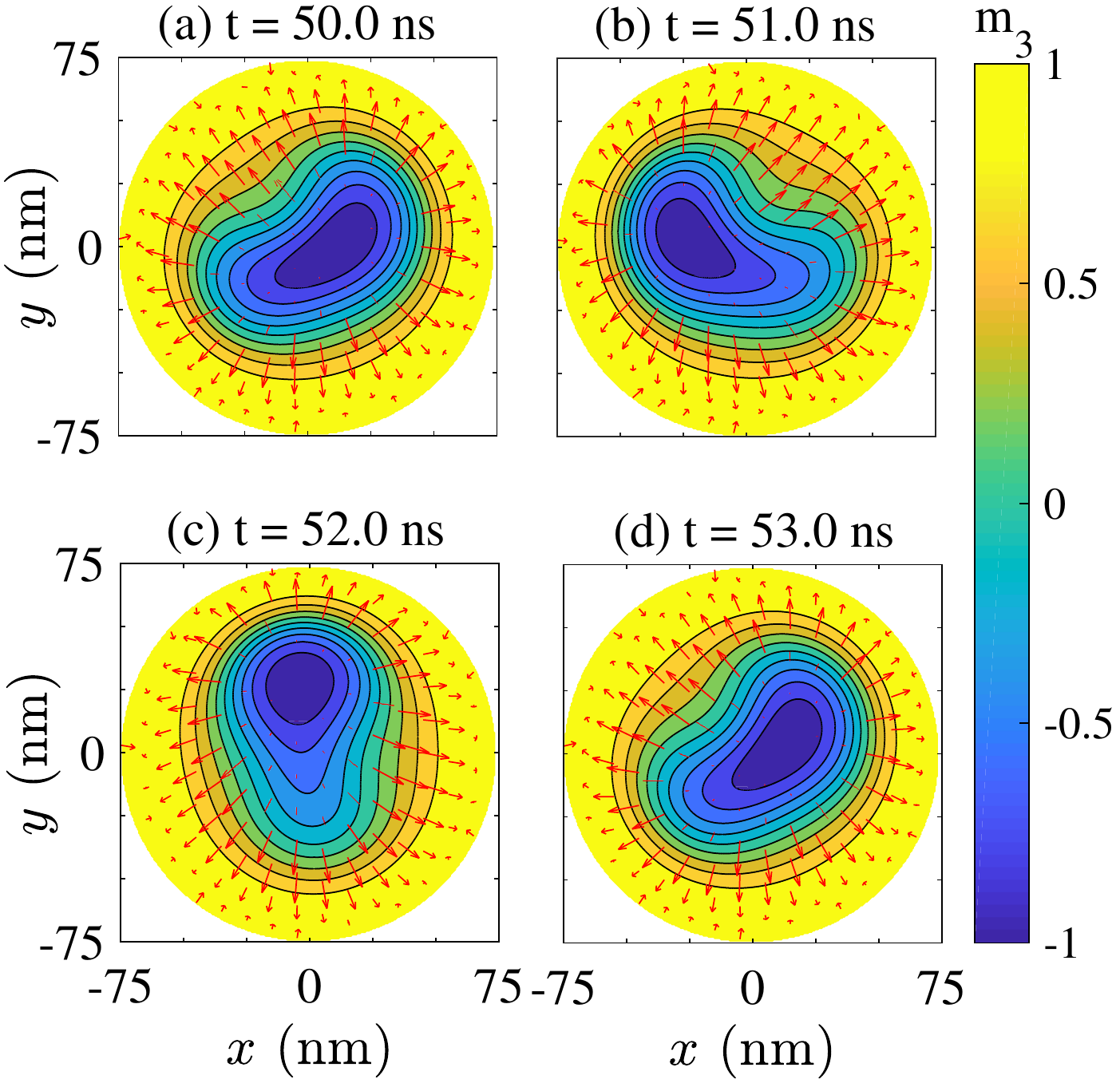}
\caption{Four snapshots of the skyrmion configuration when this is in steady-state rotation (clockwise).
We give a vector plot of $(m_1, m_2)$ and, on top of that, a contour plot for the magnetization component $m_3$ with a color code.
The time instants $t$ are shown on the corresponding entries.
}
\label{fig:rotatingSkyrmion_vectorPlot}
\end{center}
\end{figure}

Figure ~\ref{fig:rotatingSkyrmion_vectorPlot} shows four snapshots of the rotating skyrmion configuration at times when steady-state rotation has already been established.
The axially symmetric configuration of the initial (static) skyrmion is significantly distorted.
The skyrmion is not rotating in a rigid way, but the configuration is changing significantly during rotation. 
It is elongated roughly in the direction of motion.
The complete motion of the skyrmion is shown in the Supplementary Video~\cite{supp2}.

Some of the general features of the motion can be considered plausible. 
The skyrmion is driven to the left part of the nanodisc due to the spin-torque [see Fig.~\ref{fig:rotatingSkyrmion_vectorPlot}(b)].
Once the skyrmion approaches the boundary it is driven along it due to forces from the boundary, mainly the magnetostatic field
[see Fig.~\ref{fig:rotatingSkyrmion_vectorPlot}(c)].
On the other hand, the part of the trajectory when the skyrmion moves away from the boundary [Fig.~\ref{fig:rotatingSkyrmion_vectorPlot}(d)] and closes the loop, in order to start over again the cycling motion, cannot be attributed in an obvious way either to the forces from the boundary or the spin-torque.
Thus, the rotational skyrmion dynamics under the present simple set-up cannot be anticipated on the basis of any obvious arguments.
In Sec.~\ref{sec:theory} we will proceed to a detailed analysis of the skyrmion dynamics in order to account for the establishment of the auto-oscillations.

We conclude this section discussing the robustness of the phenomenon with respect to parameter values.
In order to give quantitative results (frequency and amplitude of oscillations) we choose to follow the total magnetization in the $x$ direction
\begin{equation}  \label{eq:tmagn1}
\tmagn_1 = \frac{\int m_1\, dxdy}{\int dx dy}.
\end{equation}
Since the reference layer in our simulations is chosen to be magnetized in the $x-$direction, the magnetoresistance in a MTJ devices will depend on $\tmagn_1$. 
The output power of the MTJ device would also depend on the same quantity.
We take the Fourier transform of $\tmagn_1$ and then locate the frequency for the main peak.
The Fourier transform is applied for simulation times $50-100\,\textrm{ns}$ so that the initial transient dynamics is excluded.

\begin{figure}[t]
\begin{center}
\includegraphics[width=\columnwidth]{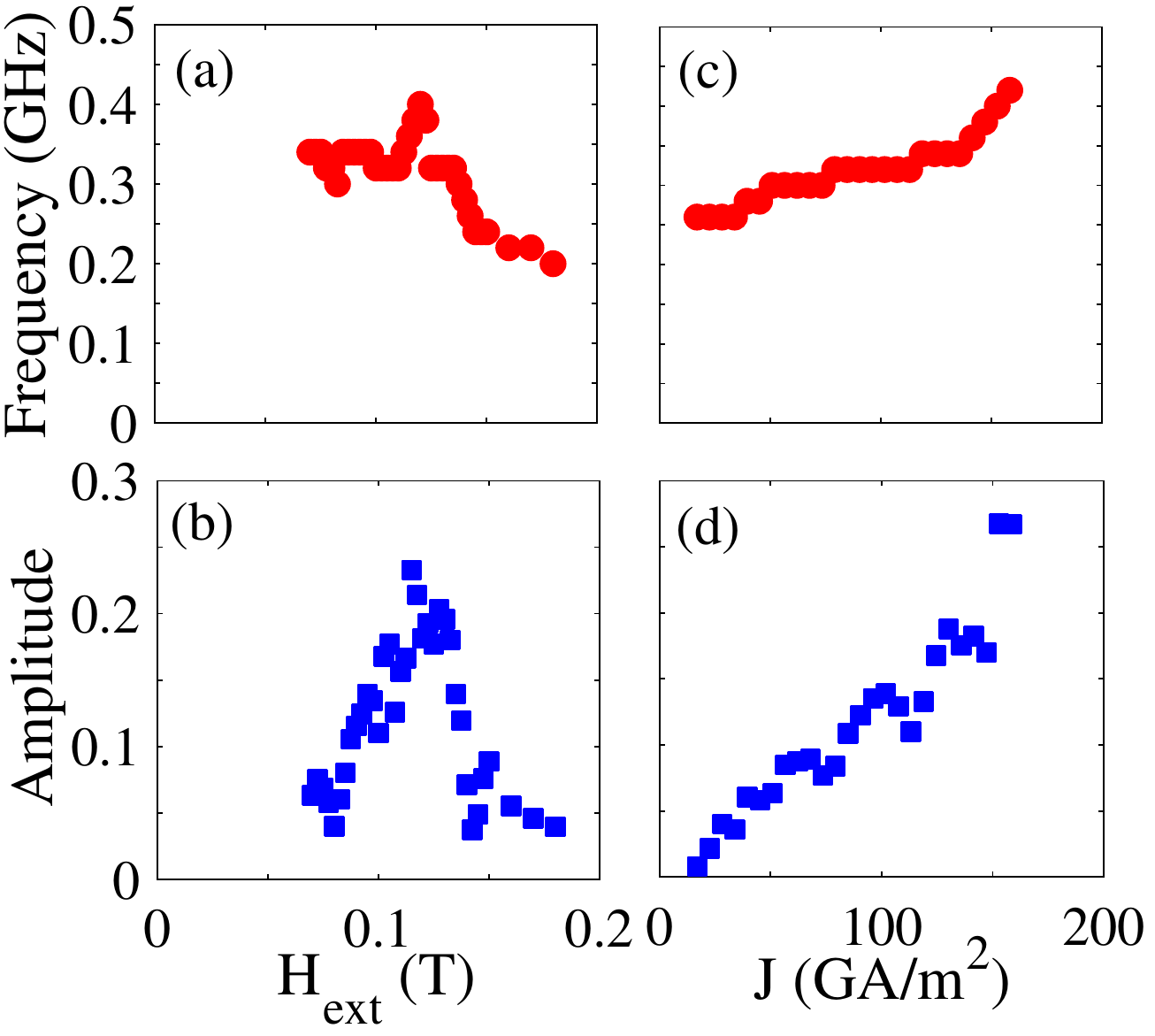}
\caption{(a) Frequency and (b) amplitude of oscillations for $\tmagn_1$ as functions of external field $\Hext$ for the parameter values in Table~\ref{table:mat_parameter} and current density $\Je=113~\textrm{GA/m}^2$.
(c) Frequency and (d) amplitude of oscillations for $\tmagn_1$ as functions of current density for external field $\Hext=0.1~{\rm T}$.
Every symbol on the graphs correspond to a separate numerical simulation.
}
\label{fig:FRQ_FIELD_MAP}
\end{center}
\end{figure}

Skyrmion rotational motion is achieved for all values of $\DM = 1.5 - 1.8\,{\rm mJ/m^2}$ for which a skyrmion is obtained as described in Sec.~\ref{sec:statcSkyrmion}.
For the set of parameter values in Table~\ref{table:mat_parameter} and for current density $\Je=113\,{\rm GA/m^2}$, we obtain skyrmion rotation for values of the external field $0.07\,{\rm T} \leq \Hext \leq 0.18\,{\rm T}$.
Fig.~\ref{fig:FRQ_FIELD_MAP}(a) shows the frequency and Fig.~\ref{fig:FRQ_FIELD_MAP}(b) shows the amplitude of magnetization oscillations as a function of external field.
If we change the DM parameter to $\DM=1.5\,{\rm mJ/m^2}$
skyrmion rotation is obtained for external fields $0.1\,{\rm T} \leq \Hext \leq 0.2\,{\rm T}$.
(We have not simulated the system for $\Hext > 0.2\,{\rm T}$.)
For the set of parameter values in Table~\ref{table:mat_parameter} and $\Hext=0.1\,{\rm T}$ we have rotational dynamics for current density in the range $\Je=20-160\,{\rm GA/m^2}$.
Fig.~\ref{fig:FRQ_FIELD_MAP}(c) shows the frequency and Fig.~\ref{fig:FRQ_FIELD_MAP}(d) shows the amplitude of magnetization oscillations as a function of current density.
We have also found sustained skyrmion rotation for disc with diameter $130-150\,{\rm nm}$.
Though optimum values of fields for skyrmion rotation changes with the disk diameter, the phenomenon is found robust with respect to changing the disc size.

We finally note that some sets of parameter values lead to oscillations of the magnetization that are more complicated than steady-state skyrmion rotation.
The results reported here do not exhaust the full range of oscillatory dynamics exhibited by the system.
While the detailed investigation of this is outside the scope of the present article, we note that many sets of parameters values lead to the skyrmion getting displaced from the center by the spin torque and it eventually relaxes at a new position off the disc center exhibiting no further dynamics.

\section{Rotation and breathing}
\label{sec:theory}

The theoretical analysis is based on the Landau-Lifshitz-Gilbert-Slonczewski (LLGS) equation for the dynamics of the normalized magnetization $\magn=\magn(x,y,t)$, whose standard form reads
\begin{equation}  \label{eq:LLGS0}
\frac{\p \magn}{\p t} = -\gamma \magn\times\bm{F} + \alpha\,\magn\times \frac{\p \magn}{\p t}
 -\STorque\,\magn\times(\magn\times\ex)
\end{equation}
with $\gamma$ the gyromagnetic ratio.
It includes Gilbert damping with parameter $\alpha$ and a spin-torque term with parameter $\STorque$.
Spin-current polarization is taken to be in-plane along the unit vector $\ex$.
The effective field $\bm{F}$ is
\begin{align}
\bm{F} = & \frac{2A}{M_s} \Delta\magn + \frac{2K}{M_s} m_3 \ez  \\
 & + \frac{2\DM}{M_s} \left[ \bnabla m_3 - \left(\bnabla\cdot\magn\right)\ez \right] + \mu_0 (\bHm + \bHext) \notag
\end{align}
where $\bHm$ is the magnetostatic field,
$\Delta$ denotes the 2D Laplacian, $\ez$ is the unit vector perpendicular to the plane, and material parameters as explained in the caption of Table~\ref{table:mat_parameter}.
For a current density $\Je$ the spin-torque parameter is
\[
\STorque = \frac{\gamma\hbar\,b}{M_s\,|e|\,d_f} \,\Je
\]
where $b$ is the polarization efficiency and $|e|$ is the electronic charge.
The LLGS Eq.~\eqref{eq:LLGS0} was used in the numerical simulations in the previous sections.

We will work with the rationalized form of the LLGS,
\begin{equation} \label{eq:LLGS}
 \p_\tau\bm{m} = -\bm{m}\times\bm{f}
 -\storque\,\bm{m}\times(\bm{m}\times\ex)
\end{equation}
where we have dropped the damping term and the effective field is
\begin{equation} \label{eq:effectiveField}
 \bm{f} = \Delta\bm{m} + \qfactor\, m_3 \ez + 2\dm\,\left[ \bnabla m_3 - (\bnabla\cdot \bm{m})\ez \right] + \bhm + \bhext.
\end{equation}
Distance is measured in exchange length units $\lex = \sqrt{2A/(\mu_0 M_s^2)}$, and we use the dimensionless time variable $\tau = t/\tau_0$, where $\tau_0=1/(\gamma \mu_0 M_s)$.
The magnetic fields $\bhm, \bhext$ are normalized to the saturation magnetization.
We have defined the dimensionless DMI parameter $\dm = \lex/\ldm$, with $\ldm = 2A/\DM$ a unit of length due to exchange and DMI.
The dimensionless spin-torque parameter is 
\begin{equation}  \label{eq:STorque}
\storque = \frac{\Je}{\Jp},\qquad \Jp = \frac{\mu_0 M_s^2 |e| d_f}{\hbar b}.
\end{equation}

The parameter values in Table~\ref{table:mat_parameter} give the length and time scales $\lex=5.0\,\textrm{nm},\,\ldm=12.9\,\textrm{nm},\,\tau_0=5.4\,\textrm{ps}$
and the dimensionless parameters $\qfactor = 0.68,\;\dm = 0.39$.    
For the spin-torque term we have $\Jp=1.24\times 10^{13}\,{\rm A/m^2}$, where we use $b=0.54$, and find $\storque = 0.0091$.

A general picture of skyrmion motion can be obtained by an approach similar to Thiele~\cite{thiele1973steady}.
We assume a magnetic configuration propagating with a constant velocity $\bm{\vel}=(\vel_1, \vel_2)$, and Eq.~\eqref{eq:LLGS} becomes
\begin{equation}  \label{eq:LLGS_travelingAnsatz}
\vel_\kappa \p_\kappa\bm{m} = \bm{m}\times\bm{f} + \storque\,\bm{m}\times(\bm{m}\times\ex).
\end{equation}
The approximation is correct to order $O(\alpha)$ and it will give the main effects of the spin-torque, considering the effect of damping as a next order phenomenon. 
We take the cross product from the right with $\p_\mu\bm{m}$ in Eq.~\eqref{eq:LLGS_travelingAnsatz} and subsequently take the scalar product with $\bm{m}$.
We then integrate every term over the plane and obtain
\begin{equation}  \label{eq:virial1}
\vel_1 = \frac{1}{\Skyrmion} (\storque T_2 + C_2),\qquad
\vel_2 = -\frac{1}{\Skyrmion} (\storque T_1 + C_1)
\end{equation}
where $\Skyrmion$ is the skyrmion number given in Eq.~\eqref{eq:skyrmionNumber}, and we have defined
\begin{equation}  \label{eq:TC}
\begin{split}
T_\mu & = \frac{1}{4\pi} \int  (\p_\mu \bm{m}\times\bm{m}) \cdot\ex\, dxdy,  \\
C_\mu & = \frac{1}{4\pi} \int (\bfext\cdot \p_\mu \bm{m})\,dxdy
\end{split}
\end{equation}
where the integration extends over the whole plane.
The vector $\bfext$ is the part of the effective field $\bm{f}$ that contains the external forces.
All contributions from terms $\bm{f}\cdot\p_\mu\bm{m}$ in $C_\mu$ that are due to forces invariant with respect to space translations, such as the internal forces of exchange, anisotropy, DMI, and a uniform external field, vanish upon integration \cite{PapanicolaouTomaras_NPB1991}.
Details of the derivation are provided in App.~\ref{Sec:analyticalFormulation}.

Eqs.~\eqref{eq:virial1} have been obtained for a two-dimensional system extending to spatial infinity.
For the confined geometry in our numerical simulations, we will use them as an approximation.
In order for the approximation to be valid the magnetostatic field due to the boundary of the magnetic element is treated as an external force and it gives the $\bfext$.

\begin{figure}[t]
\begin{center}
\includegraphics[width=\columnwidth]{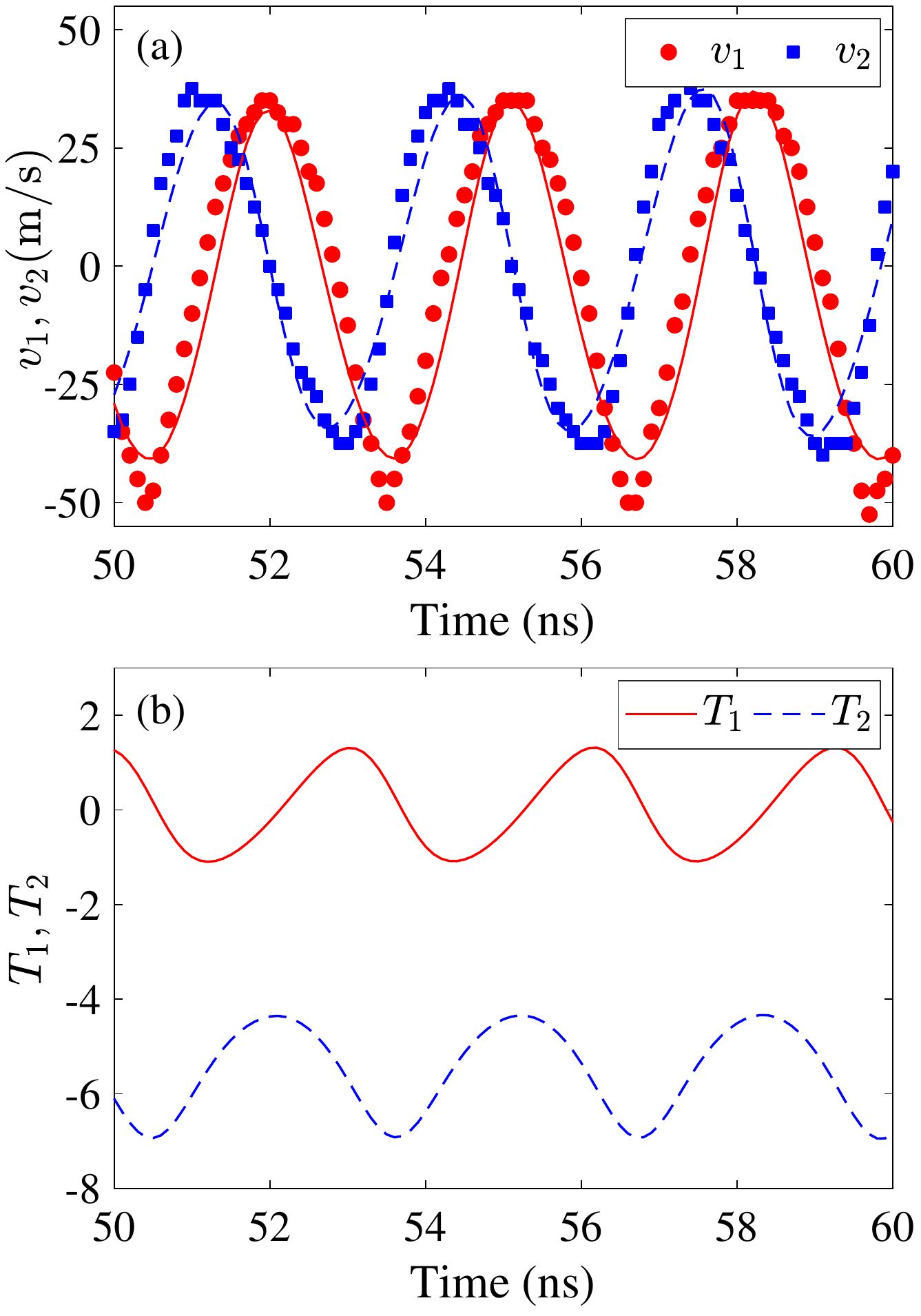}
\caption{(a) The velocity components $(\vel_1, \vel_2)$ of the skyrmion as functions of time, found in the simulation, for a few cycles after the steady state has been established (red and blue symbols, respectively).
We also plot the corresponding right hand sides of Eq.~\eqref{eq:virial1}, in red-solid (for $\vel_1$) and blue-dashed (for $\vel_2$) lines, 
showing that the formulas are indeed capturing the skyrmion dynamics.
The period of rotation is approximately $3.1\,\textrm{ns}$ (or $574\,\tau_0$).
(b) The quantities $T_1, T_2$ of Eq.~\eqref{eq:TC} as functions of time.
They are oscillating in phase with the corresponding velocity components $-\vel_2, \vel_1$.
}
\label{fig:velocityVStime}
\end{center}
\end{figure}

We calculate the skyrmion velocity in the simulation using finite differences between successive positions of the skyrmion.
The velocity components are plotted in Fig.~\ref{fig:velocityVStime}(a) for a period of a few cycles after the steady rotation has been established.
Both velocity components are oscillating with time.
Furthermore, we calculate the quantities in Eq.~\eqref{eq:TC} during the simulation using the numerically found magnetization configuration, and we then substitute these values in Eq.~\eqref{eq:virial1} in order to obtain the predicted velocity.
We restore physical units in Eqs.~\eqref{eq:virial1} multiplying by the unit of velocity
$\vel_0 = \lex/\tau_0 = 920\,\textrm{m/sec}$.
In Fig.~\ref{fig:velocityVStime}(a) we plot the velocity components given by Eq.~\eqref{eq:virial1} and we find excellent agreement with the numerical simulation results.
This justifies the approximations performed in order to obtain Eq.~\eqref{eq:virial1}. 

Let us follow in detail the prediction of Eq.~\eqref{eq:virial1} during the skyrmion rotation.
The quantities $C_{1,2}$ are due to forces from the boundaries of the nanodisc, and they drive the skyrmion along the boundaries.
On the other hand, the quantities $T_{1,2}$ are central in
driving the skyrmion motion, but their time dependence cannot be predicted in any obvious way. 
We plot the numerically calculated $T_{1,2}$ in Fig.~\ref{fig:velocityVStime}(b) and we observe that $T_2$ and $-T_1$ are oscillating in phase with the corresponding velocity components $\vel_1, \vel_2$.
This is in agreement with the rotational motion of the skyrmion and Eq.~\eqref{eq:virial1}.
The quantities $C_{1,2}$ (not shown in the figure) also oscillate as the skyrmion is rotating, as a simple consequence of the fact that the skyrmion periodically approaches and goes away from the boundary (which is the main source of the magnetostatic field).

\begin{figure}[t]
\begin{center}
\includegraphics[width=\columnwidth]{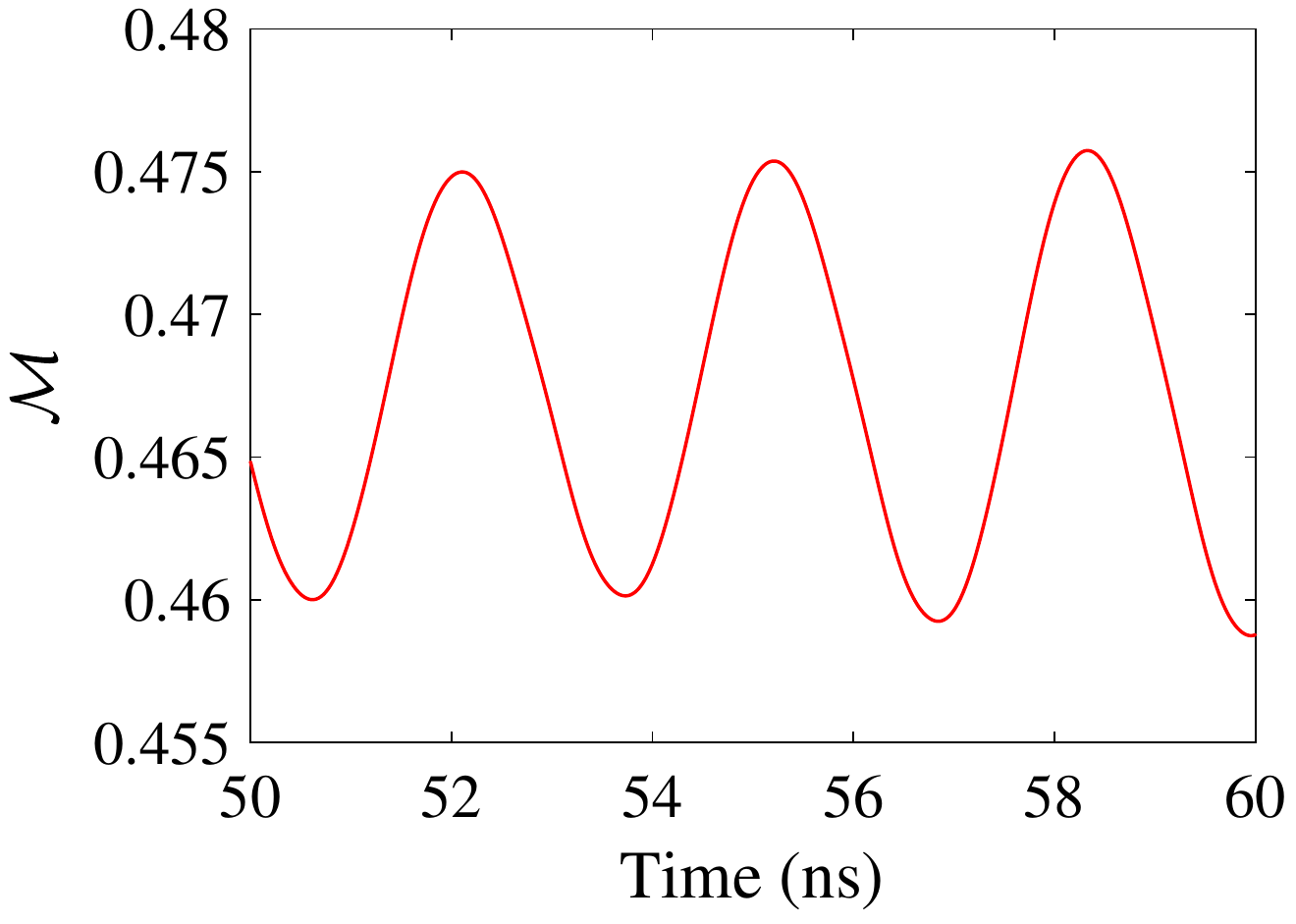}
\caption{(a) Total magnetization $\tmagn$, in Eq.~\eqref{eq:tmagn}, as function of time.
The period of oscillations is the same as for the velocity components in Fig.~\ref{fig:velocityVStime}, that is, the skyrmion is breathing at the same frequency as it is rotating.
}
\label{fig:tmagn}
\end{center}
\end{figure}

For an explanation of the achievement of a steady state rotation for the skyrmion, it is crucial to discuss the phenomenon that produces the oscillations of $T_{1,2}$.
We will see that this is linked to an eigenmode of the system and it is, thus, instructive to consider the conservative model, that is, Eq.~\eqref{eq:LLGS} for $\alpha=0,\, \storque=0$.
The key point is to follow the normalized total magnetization out of the plane:
\begin{equation}  \label{eq:tmagn}
\tmagn = \frac{\int m_3\, dx dy}{\int dx dy},
\end{equation}
which is actually a quantity typically measured in experiments.
The total magnetization $\tmagn$ would be a conserved quantity in a typical model where only isotropic exchange and uniaxial anisotropy interactions are present.
But, the presence of DMI in Eq.~\eqref{eq:effectiveField} breaks the symmetry of rotations around the third magnetization axis ($\ez$).
As a consequence, $\tmagn$ is not a conserved quantity, and thus, oscillations of $\tmagn$ are allowed. This means that breathing of the skyrmion may be expected, and it is indeed an eigenmode of the system \cite{SchuetteGarst_PRB2014}.
In Fig.~\ref{fig:tmagn} we plot $\tmagn$ for the rotating skyrmion of our simulations and show that this is oscillating.
Note that the frequency of oscillations of $\tmagn$ is the same as the frequency of rotation of the skyrmion, and this is very different than the eigenfrequency of a freely breathing skyrmion (see App.~\ref{sec:breathingRotation}).

Breathing oscillations for an axially symmetric skyrmion, where the skyrmion radius is periodically varying, have been studied in Refs.~\cite{mochizuki2012spin,kim2014breathing}.
But, it should be stressed that in both these references, breathing is driven by an ac field and dissipative effects.
We will explain that the present work implicates the breathing mode in a different way: it uses uniform fields (spin-torque) and exploits symmetry breaking in order to generate emerging oscillations that will drive the periodic motion.

In order to develop our arguments we consider an axially symmetric skyrmion configuration expressed in the spherical parametrization of the magnetization via the angles
\begin{equation}  \label{eq:m_rhophiz}
\Theta = \Theta(\rho),\qquad \Phi = \phi + \phi_0
\end{equation}
where $\rho, \phi$ are polar coordinates and $\phi_0$ is a 
phase angle that is often called the chirality.
The
constant
values $\phi_0=\pm\pi/2$ give Bloch skyrmions while $\phi_0=0,\pi$ give N\'eel skyrmions.
For our interfacial DMI the value $\phi_0=0$ is energetically preferable.
We substitute Eq.~\eqref{eq:m_rhophiz} in the definition \eqref{eq:TC} for $T_{1,2}$ and obtain
\begin{equation}  \label{eq:T1T2AxialSymmetry}
\begin{aligned}
T_1 = \sin\phi_0\, T,\quad
T_2 = \cos\phi_0\, T,  \qquad \\
T = \frac{1}{2} \int \left[ \p_\rho\Theta + \frac{\sin(2\Theta)}{2\rho}  \right] (2\pi\rho\, d\rho).
\end{aligned}
\end{equation}

Let us assume an oscillation of the skyrmion size (radius), i.e., breathing of the skyrmion. This is obviously expected to give rise to an oscillation of $\tmagn$ (such as shown in Fig.~\ref{fig:tmagn}) as well as $T$.
We may write
\begin{equation}  \label{eq:TOsc}
\begin{gathered}
T \approx T_0 + \epsilon T' \cos(\omega t)
\end{gathered}
\end{equation}
where $T_0$ is the mean value of $T$, the product
$\epsilon T'$ gives the amplitude of oscillations with angular frequency $\omega$.
The formula is valid when the parameter $\epsilon$ is small.

Furthermore, an oscillation of $\tmagn$ should be accompanied by an oscillation of the phase $\Phi$ (equivalently, of $\phi_0$) of the skyrmion,
since $m_3$ and $\Phi$ are conjugate variables in the Hamiltonian formulation of the Landau-Lifshitz equation. 
Such oscillations are observed in our simulations.
This phenomenon has not been mentioned in any of the reports on the breathing mode, to the best of our knowledge.
For an axially symmetric skyrmion the phase angle does not depend on the space variable and we write
\begin{equation}  \label{eq:phi0Osc}
\phi_0 \approx \epsilon \sin(\omega t).
\end{equation}
The phase angle $\phi_0$ in Eq.~\eqref{eq:phi0Osc} and the quantity $T$ in Eq.~\eqref{eq:TOsc} (that oscillates in phase with $\tmagn$) 
are taken to oscillate with the same angular frequency and at a phase difference of $\pi/2$ in the approximation that the system behaves as a harmonic oscillator for the
conjugate
variables $\phi_0$ and $\tmagn$.
This is valid in the linear approximation, that is, for small oscillation amplitudes.

We substitute Eqs.~\eqref{eq:TOsc} and \eqref{eq:phi0Osc} in the definition \eqref{eq:T1T2AxialSymmetry} 
and keep terms up to order $O(\epsilon)$ to obtain
\begin{equation}  \label{eq:T_oscillations}
T_1 \approx \epsilon T_0\,\sin(\omega t),\qquad
T_2 \approx T_0 + \epsilon T'\,\cos(\omega t).
\end{equation}
Formulas \eqref{eq:T_oscillations} are in agreement with the numerical result plotted in Fig.~\ref{fig:velocityVStime}(b).
We conclude that breathing of the skyrmion gives rise to oscillations of $T_{1,2}$ and the latter drive the skyrmion motion according to Eq.~\eqref{eq:virial1}.
The constant term in $T_2$ in Eq.~\eqref{eq:T_oscillations} is balanced by the force from the disc boundary when the skyrmion is not at the disc center.
The result is a periodic rotational motion of the skyrmion centered off the disc center.

The main arguments that we have used in our explanation of the sustained skyrmion rotation do not include dissipative effects in the Landau-Lifshitz equation.
Specifically, Eqs.~\eqref{eq:virial1} are correct for $\alpha=0$, and breathing is an internal mode of the conservative system.
Despite that damping is not the main effect driving the skyrmion motion, it plays a role in the achievement of stable rotational motion.
In order to obtain auto-oscillations the internal (breathing) mode and the externally (spin-torque) induced motion should be synchronized.
This is achieved in the simulations thanks to a balance between damping and spin-torque.
The amplitude of oscillations is tuned by these effects such that energy loss and gain are balanced and also the coupling of frequencies is achieved.
As a result, the rotational motion is stabilized and sustained.
Our theoretical arguments cannot predict the parameter values for which the above mentioned energy balance and mode coupling is achieved.

In our simulations the breathing (oscillation of $\tmagn$ in Fig.\ref{fig:tmagn}) and rotational modes (oscillation of $X$, $Y$ in Fig.\ref{fig:XY}(a)) are excited initially with different periods, and they are gradually converging to the same period confirming the existence of a coupling between them.
We give some further details for the two modes in App.~\ref{sec:breathingRotation}.

\section{Concluding remarks}

We have shown the existence of auto-oscillations for a skyrmion in a nanodisc by numerical simulations, and we have explained by theoretical arguments that skyrmion auto-oscillations are obtained thanks to the DMI.
The chiral symmetry-breaking is crucial for the achievement of skyrmion auto-oscillations under constant in time and uniform in space spin-torque.
It allows for skyrmion breathing together with a periodic variation of the in-plane component of the magnetization.
The result is an emerging periodic force on the skyrmion due to spin-torque that gives sustained rotational skyrmion motion.
Our theory captures the principle of sustained skyrmion rotation and is based on a model for an axially symmetric skyrmion.
On the other hand, numerical simulations show a more complicated and changing skyrmion profile as this is rotating.

The phenomenon is very robust with respect to changing parameter values.
More complicated oscillatory dynamics has also been numerically found.
We have not explored the full range of oscillatory dynamics exhibited by the system.
A full phase diagram could be produced by expanding the results of the present paper.

For completeness we note that the theoretical arguments of our paper could be repeated and applied when assuming a magnetostatic field (instead of, or in addition to DMI).
Thus, auto-oscillations of magnetic bubbles may be expected due to the always-present magnetostatic field.
But, it appears that the magnetostatic field is too weak for the phenomenon to be achievable.
In addition, it is more complicated to obtain skyrmions (or bubbles) when the DMI is not present.

\acknowledgments
This work was initiated at the Max-Planck Institute for the Physics of Complex Systems in Dresden, during the Workshop "Topological Patterns and Dynamics in Magnetic Elements and in Condensed Matter".
Partial support by the Department of Science and Technology, India, under nanomission program is gratefully acknowledged. The authors thank the IIT Delhi HPC facility for computational resources. N.~S. acknowledges support from the Ministry of Human Resource Development (MHRD), India.

\appendix

\section{Derivation of Thiele's equation}
\label{Sec:analyticalFormulation}

In our analytical calculations we consider a steady-state configuration with a constant velocity $\bm{\vel}=(\vel_1,\vel_2)$, that is, we consider the traveling wave ansatz
\begin{equation}  \label{eq:2}
\magn = \magn(x - \vel_1\tau, y - \vel_2\tau).
\end{equation}
We have $\p_\tau\bm{m}=-\vel_\kappa\p_\kappa\bm{m}$, which is substituted in Eq.~\eqref{eq:LLGS} to give Eq.~\eqref{eq:LLGS_travelingAnsatz} where we neglect damping (set $\alpha=0$).
Taking the cross product in Eq.~\eqref{eq:LLGS_travelingAnsatz} from the right with $\p_\mu \bm{m}$ and then a scalar product with $\bm{m}$ we obtain
\begin{equation}  \label{eq:virialExplicit}
\vel_\kappa \epsilon_{\mu\kappa}\skyrmion=-\bm{f}\cdot\p_\mu\bm{m}+\storque(\bm{m}\times\p_\mu\bm{m})\cdot\ex.
\end{equation}
A fundamental result is that the first term on the right-hand side can be written as the total divergence of a tensor $\sigma_{\mu \nu}$ \cite{PapanicolaouTomaras_NPB1991}
\begin{equation}  \label{eq:energyDerivative}
-\bm{f}\cdot\p_\nu\bm{m}=\p_\lambda\sigma_{\nu\lambda}.
\end{equation}
This is valid in the absence of forces with explicit space dependence.
If an external non-uniform field $\bm{f}_{ext}$ is present, then we may make the replacement $\bm{f} \rightarrow \bm{f} + \bm{f}_{ext}$ and relation \eqref{eq:energyDerivative} becomes
\begin{equation}  \label{eq:externalForce}
-\bm{f}\cdot\p_\nu\bm{m}=\p_\lambda\sigma_{\nu\lambda}-\bm{f}_{ext}\cdot\p_\nu\bm{m}.
\end{equation}

We now return to Eq. \eqref{eq:virialExplicit}, substitute Eq.~\eqref{eq:externalForce}, and integrate over all space.
The term that is a divergence (and contains the conservative forces) vanishes upon integration over all space and we finally obtain Eq.~\eqref{eq:virial1}.

\section{Breathing and Rotation without STT}
\label{sec:breathingRotation}

We simulate the breathing and rotational modes separately in order to show that they are, in general, uncoupled and have different oscillation frequencies.
No spin-torque was applied in the simulations of this section.

\begin{figure}[h]
\begin{center}
\includegraphics[width=\columnwidth]{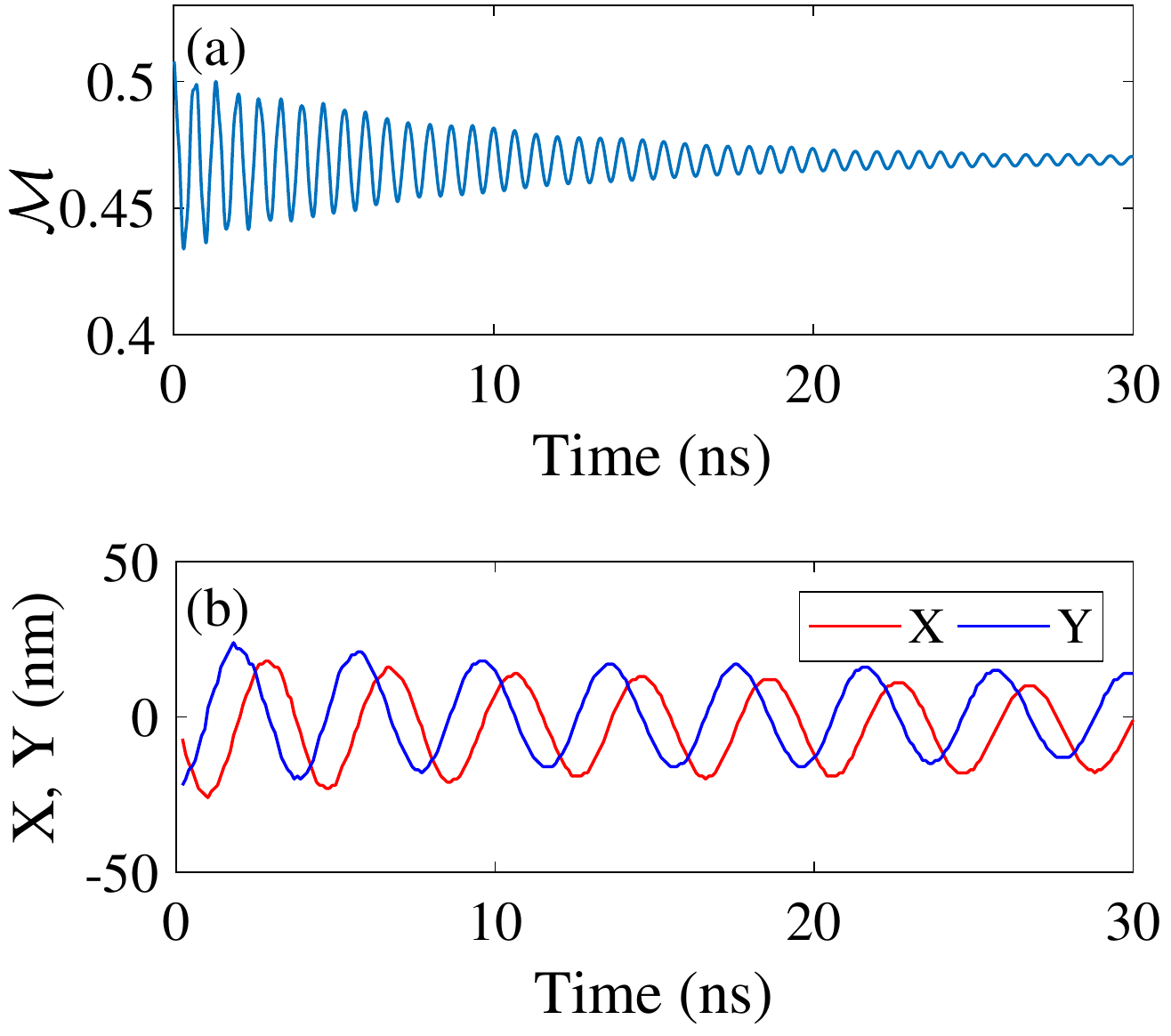} 
\caption{(a) The skyrmion is initially perturbed and the breathing mode is excited. We show the total magnetization $\tmagn$ as a function of time. The period of the breathing mode is approximately $0.68\,{\rm ns}$.
(b) The skyrmion is initially displaced from the center of the nanodisc.
We show the coordinates $X$ and $Y$ of the skyrmion position as functions of time, as the skyrmion relaxes back in a spiral trajectory to the centre of the nanodisc.
The period of the rotation is approximately $4\,{\rm ns}$.
}
\label{fig:breathingRotation}
\end{center}
\end{figure}

In order to simulate the breathing mode, we start with the skyrmion shown in Fig.~\ref{fig:staticSkyrmion} and apply an additional external field 0.01~T (this is on top of the 0.1~T field for initially obtaining the skyrmion).
The system is relaxed and the extra 0.01~T field is removed at $t=0~{\rm ns}$ and we further run the simulation using damping $\alpha=0.01$.
Fig.~\ref{fig:breathingRotation}(a) shows the total magnetization $\tmagn$ which is oscillating with a period of approximately $0.68\,{\rm ns}$ demonstrating the breathing skyrmion mode.

In order to simulate rotation of skyrmion around the disc center, we start with the skyrmion shown in Fig.~\ref{fig:staticSkyrmion} and apply a 0.01~T in-plane field so that it is pushed slightly off the disc center.
The extra in-plane field is removed at $t=0~{\rm ns}$ and we further run the simulation using damping $\alpha=0.01$.
Fig.~\ref{fig:breathingRotation}(b) shows the coordinates of the skyrmion position as the skyrmion moves in a spiral trajectory to the centre of the nanodisc.
The period of the rotation is approximately $4\,{\rm ns}$ and this is quite different than the period of the breathing mode seen earlier.
Comparing Figs.~\ref{fig:breathingRotation}(a) and (b), we conclude that in the absence of STT, breathing and rotation of the skyrmion are uncoupled.

\bibliography{refs}
\bibliographystyle{aipnum4-1}

\end{document}